\documentclass[trackchanges,twocolumn]{aastex701}
\usepackage{amsmath}
\usepackage{subcaption}
\begin{document}

\title{X-ray Polarization Signatures from Comptonization by Magnetic Reconnection Plasmoids}

\author[orcid=0000-0002-7054-9053]{John Groger}
\affiliation{Physics Department and the McDonnell Center for the Space Sciences, Washington University in St. Louis, 1 Brookings Drive, Saint Louis, MO 63130, USA}
\email[show]{groger.j@wustl.edu}

\author[orcid=0000-0002-9705-7948]{Kun Hu}
\affiliation{Physics Department and the McDonnell Center for the Space Sciences, Washington University in St. Louis, 1 Brookings Drive, Saint Louis, MO 63130, USA}
\email[show]{hkun@wustl.edu}

\author[orcid=0000-0002-1084-6507]{Henric Krawczynski} 
\affiliation{ Physics Department, the McDonnell Center for the Space Sciences, and the Center for Quantum Leaps, Washington University in St. Louis, 1 Brookings Drive,
Saint Louis, MO 63130, USA}
\email[show]{krawcz@wustl.edu}

\begin{abstract}

Emission from X-ray binaries in the hard spectral state is dominated by high-energy radiation attributed to the Compton scattering of seed photons. The prevalent model of the Comptonization by hot electrons or pairs faces the problem of rapid radiative cooling of the emitting particles. A proposed alternative mechanism is the Comptonization by scattering off fast plasmoids formed during magnetic reconnection. 
In this work, we simulate a simplified model of the plasmoid chain with Monte Carlo radiation transport and report on spectropolarimetric properties. 
We find that the Comptonization by transrelativistic bulk plasmoids is not only able to reproduce the 100\,keV spectral cutoff, 
but furthermore produces X-rays that are above 1\,keV that are strongly polarized perpendicular to the reconnection layer. The polarization is stronger than that from the Comptonization by an isotropic hot plasma owing to the confinement of the motion of the scattering plasmoids in the plane of the reconnection layer.   
The dependence of polarization on azimuthal viewing angle is discussed, along with possible locations for the plasmoid chain in an equatorial current sheet or the sheath of the black hole's relativistic jet.

\end{abstract}



\section{Introduction} 
The {\it IXPE} satellite launched in 2021 December has opened up the new field of X-ray polarimetry. The observations can constrain the geometry of the emission regions as X-rays easily scatter and scattering gives rise to strong polarization perpendicular to the scattering plane. The observations furthermore constrain the magnetic fields in the sources as (i) synchrotron emission is strongly polarized perpendicular to the magnetic field in the plane of the sky \citep{1965ARA&A...3..297G}, and (ii) Faraday rotation can depolarize the X-ray emission \citep{2009ApJ...703..569D, 2024ApJ...977..201B, 2026ApJ...998L..47K}. The X-ray emission from black holes in the hard spectral state is thought to originate from the Comptonization of longer-wavelength emission, probably coming from the accretion disk.
The Comptonization of disk photons and the reflection of some of the Comptonized emission off the accretion disk might be able to account for the observed energy spectra up to the thermal cutoff at 100~keV. From a theoretical standpoint, the model requires significant fine-tuning as the inverse-Compton cooling is likely to cool the electrons to temperatures well below the observed 100~keV \citep{2017ApJ...850..141B, 2025ApJ...993...54K}. Furthermore, the model must explain the high polarization degree (PD) of some sources as observed by IXPE in the 2–8 keV band, such as Cygnus X-1 at 4\% \citep{2022Sci...378..650K} and IGR J17091-3624 at 8\% \citep{2025MNRAS.541.1774E,2025ApJ...989..165D}. 
An alternative Comptonization scenario proposed by \cite{2017ApJ...850..141B} arises from relativistic magnetic reconnection, which can produce a magnetic flare consisting of a chain of plasmoids with transrelativistic velocities along the current sheet. Electrons in these plasmoids cool to nonrelativistic energies, so in contrast to thermal Comptonization, photons are scattered by the plasmoid bulk motions. This process, known as “cold-chain Comptonization,” was found to reproduce the 100~keV spectral cutoff without maintaining a hot coronal plasma, which would require a heating mechanism to counteract strong inverse-Compton cooling \citep{2017ApJ...850..141B}. Due to the anisotropic configuration of chain Comptonization, it is expected to be capable of generating highly polarized emission. 

To account for the spectrum and PD observed for Cyg X-1 in particular, coronal plasma ejected by magnetic flares has been proposed to outflow as a transrelativistic wind from the accretion disk \citep{1999ApJ...510L.123B,2023ApJ...949L..10P}. Beloborodov (\citeyear{1998ApJ...496L.105B}) found that a vertical pair wind propagating with 50\% of the speed of light away from the accretion disk can produce a polarization parallel to the disk normal, as Thomson scattering modifies the initially perpendicular polarization predicted by Chandrasekhar (\citeyear{1960ratr.book.....C}). Other models propose that such an outflow occurs along a conical or paraboloidal jet sheath \citep{2024MNRAS.528L.157D,2024Ap&SS.369...68M}. The jet sheath as a site of dissipation and candidate Comptonizing region has been characterized with resistive GRMHD simulations by \cite{2025ApJ...979..199S}. 

In this work, we adopt a simplified slab geometry and use a radiation transport code to simulate the spectropolarimetric predictions for the cold-chain Comptonization scenario in a reconnection layer. We first describe the model in Section \ref{setup}. We report on an approximate agreement with observed spectra and on polarization as functions of energy and observer inclination in Section \ref{results}. Our initial analysis is agnostic about the configuration of the layer within the black hole system. In Section \ref{discussion}, these results are then connected to the proposed hollow-cone/jet-sheath geometry, and to an equatorial current sheet, at a viewing angle of $i \sim 28^{\circ}$ for Cyg X-1.

\section{Reconnection Layer and Simulation Setup} \label{setup}
\begin{figure*}[t]
    \centering
    \includegraphics[width=1.0\textwidth]{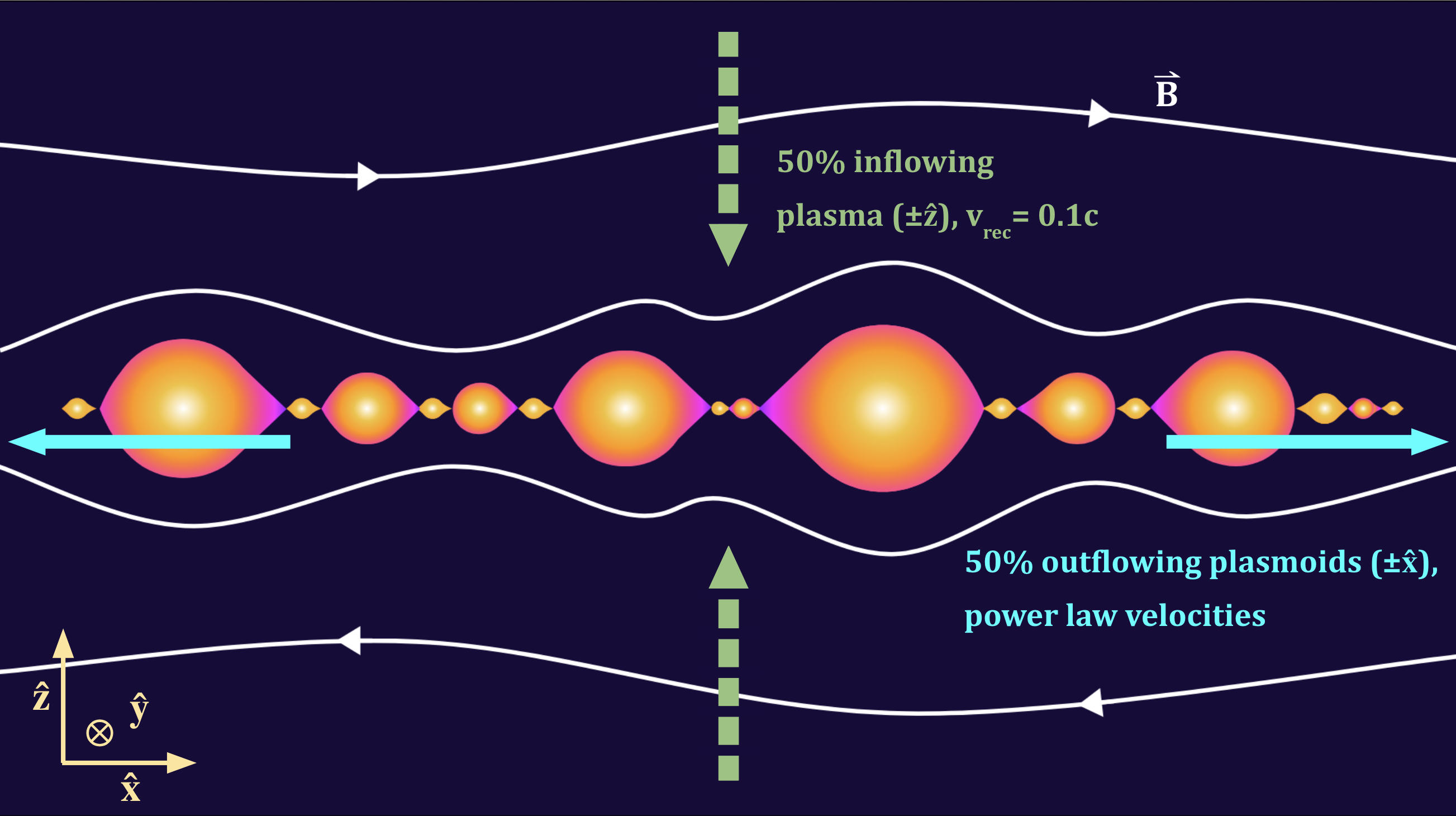}
    \caption{A schematic illustration of the reconnection layer model used in our simulations. Electrons are sampled from plasma inflowing with $v_{rec}=0.1c$ ($\pm \hat z$, green/dashed) and from plasmoids outflowing with transrelativistic velocities ($\pm \hat x$, blue/solid). In our simplified model, these regions are not spatially separated but are sampled with equal probability.}
    \label{fig:plasmoidchain}
\end{figure*}

\begin{figure}
    \centering
    \includegraphics[width=1.0\linewidth]{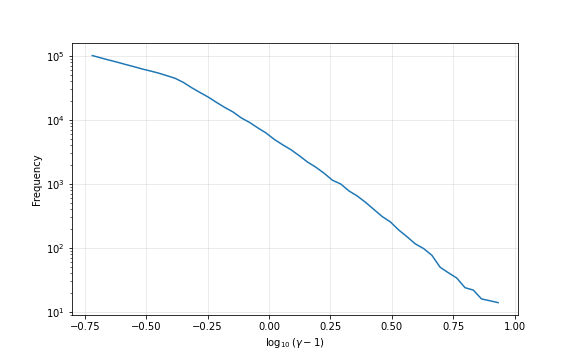}
    \caption{Distribution of plasmoid Lorentz factors in a simulation with $a_{min}=0.4$, $a_{max}=100$, sampled from the broken power-law distribution in Equation (\ref{eq:powerlaw}).
    }
    \label{fig:lfac}
\end{figure}

We simulate the Comptonization by mildly relativistic 
plasmoids or flux ropes with the simplified model that 
Beloborodov (\citeyear{2017ApJ...850..141B}) introduced. The model incorporates the main features of reconnection as derived from 2D particle-in-cell (PIC) simulations \citep[][and references therein]{2025ARA&A..63..127S}. 
The basic geometry is shown in Figure \ref{fig:plasmoidchain}. The reconnecting current sheet is in the $x-y$ plane and the plasma streams into the reconnection zone along the positive and negative $z$-direction. The plasmoids or flux ropes move out of the reconnection zone along the positive and negative $x$-axis. The electrons are assumed to be cold owing to the Compton cooling by disk photons and are uniformly distributed within a layer with a vertical optical depth of $\tau$ at the $z=0$ plane. The Comptonization proceeds via scattering off two components:
(i) from the upstream plasma moving into the reconnection zone with the velocity $\beta_{\rm rec}=0.1$ along $\pm \hat z$, and (ii) via scattering off the plasmoids moving along $\pm \hat x$ 
with Lorentz factors drawn from the distribution

\begin{equation}
P(a)\propto
\begin{cases}
a^{\psi_1}, & a < 1 \\
a^{\psi_2}, & a > 1
\end{cases}
\label{eq:powerlaw}
\end{equation}
where the velocity parameter $a=\beta^2\gamma^2$. This distribution follows from Beloborodov's model for the power deposited in the plasmoid chain, in which Compton drag leads to an inverse relation between the velocity parameter $a$ and plasmoid size. More explicitly, we adopt $\psi_1=q/2 - 2$ and $\psi_2=q-3$ from Beloborodov's calculation, where the distribution of plasmoid sizes follows $N \sim w^{-q}$. \cite{2016MNRAS.462...48S} report that $q$ is close to 0, i.e., $N\sim$ constant, based on PIC simulations, in line with the predictions of \cite{2012PhRvL.109z5002H}. Throughout this work, we adopt $\psi_1=-2$ and $\psi_2=-3$. This treatment is consistent with \cite{2018MNRAS.475.3797P} and \cite{2019MNRAS.482...65C}, who find that the Comptonized/synchrotron spectral properties of the plasmoid chain are governed primarily through the velocity distribution, with size effects largely mediated through velocity rather than acting independently. The sampling in Figure \ref{fig:lfac} shows the desired power-law break at $a=1$ ($\gamma = \sqrt2$) as set by PIC simulations and Compton drag calculations. The simulations are sensitive to the choice of optical depth as well as $a_{min}$ for the sampling, while $a_{max}$ is fixed at 100. Beloborodov reports that $a_{max}$, which scales directly with the magnetization $\sigma$, does not greatly affect results in the regime $\sigma \gg 1$.

The uniform-layer geometry provides an effective coarse-grained description of the emitting region and is therefore well suited for predicting the time-averaged spectrum and polarization. We can therefore focus on the dependence of the observable radiation on macroscopic properties, such as the geometry of the current sheet and jet sheath, as well as the viewing inclination, rather than on the detailed locations, sizes, and shapes of individual plasmoids. This choice is supported by \cite{2016MNRAS.462...48S} and \cite{2024MNRAS.531.4781Z}, who find isolated plasmoids have a modest aspect ratio ($\sim1.5:1-1.6:1$) and show analytically that this elongation alone produces only a $\sim16\%$ synchrotron PD; the larger polarization signatures they report arise from plasmoid mergers rather than shape. These macroscopic geometric dependencies may provide more robust and astrophysically informative constraints on the black hole system. One limitation of this approach is that we do not track plasmoid position relative to the z-boundaries. Since \cite{2016MNRAS.462...48S} find the largest plasmoids reach $w_{max}/L \sim 0.2$, preferential escape of photons scattered by such large, slow plasmoids near the boundary is a possible second-order effect not captured here, which would be expected to enhance flux at the lower end of the spectrum.

The Monte Carlo simulation is adapted from the radiation transport code used by \cite{2024ApJ...977L..10K}. We modify the code by adding a module to draw scatterers with velocity $\gamma_e$ for individual electrons taken from the bulk plasmoid velocity distribution above rather than a Maxwell--Boltzmann distribution for a specified plasma temperature. We set the direction $\vec{n}_e$ for each scatterer to be purely along the $x$- and $z$-axes as described above instead of being drawn isotropically. As the velocity has no $y$-component, the problem is translation invariant in y. However, the radiative transfer is fully 3D, and constraining the scatterer’s lab-frame velocity direction to $\pm \hat x$ and $\pm \hat z$ does not constrain the scattering angle in the electron rest frame, which is sampled from the full Klein--Nishina distribution after boosting.

Seed photons are injected at the midplane of the layer ($z=0$) with a blackbody spectrum, uniformly distributed in the $x-y$ plane, with polarization and angular distribution set by Chandrasekhar’s optically thick scattering atmosphere \citep{1960ratr.book.....C}. We note that this modeling choice is necessary because \cite{2017ApJ...850..141B} does not specify a seed photon polarization state. The photons escape at the upper and lower boundary of the reconnection region with a vertical optical depth of $\tau$. Our Monte Carlo radiation transport simulation tracks individual photon trajectories, photon four-wavevectors $k^\alpha$, and polarization four-vectors $f^\alpha$ \citep{2017ApJ...850...14B,2024ApJ...977L..10K}.
Scatterings are implemented by drawing a plasmoid four-velocity, transforming $k^\alpha$ and $f^\alpha$ into the reference frame of the plasmoid, transforming the statistical weight of the photon, its PD 
(a Lorentz invariant), and $f^\alpha$ into
Stokes parameters $I$, $Q$ and $U$, and 
using Fano's scattering matrix to effect the scattering off an electron (or positron)
accounting for the properties of the Klein--Nishina cross section \citep{1957RvMP...29...74F,1961RvMP...33....8M}. 
We use the rejection method to implement the suppression of the scatterings in the Klein--Nishina regime and 
dependence of the cross section on the relative velocity of the photon and the plasmoid with the help of a rejection algorithm.
Note that our treatment assumes that a single cold electron (or positron) inside the plasmoid scatters the photons, neglecting the effect of higher scattering orders. 
The treatment is justified as the first scattering order always dominates the outcome of a reflection, and our model is agnostic of the orientation of the scattering surface. Photons exiting the reconnection and scattering layer are sorted into 13 inclination angle bins with centers between $15^{\circ}.9$ and $87^{\circ}.8$ relative to the $z$-axis (i.e., the reconnection layer normal), and their Stokes $I$, $Q$, and $U$ parameters are calculated and added to the running sums of their bins. We also record these quantities for certain azimuthal angles $\phi$ measured in the $x-y$ plane. We record the Compton amplification factor $A=\bar E_{Esc} / E_S$, the ratio of injected to escaped photon energy. We note that the amplification factor required to fit the observations depends on the mean energy of the seed photons.

\section{Results} \label{results}

\begin{figure}
    \centering
    \includegraphics[width=1.0\linewidth]{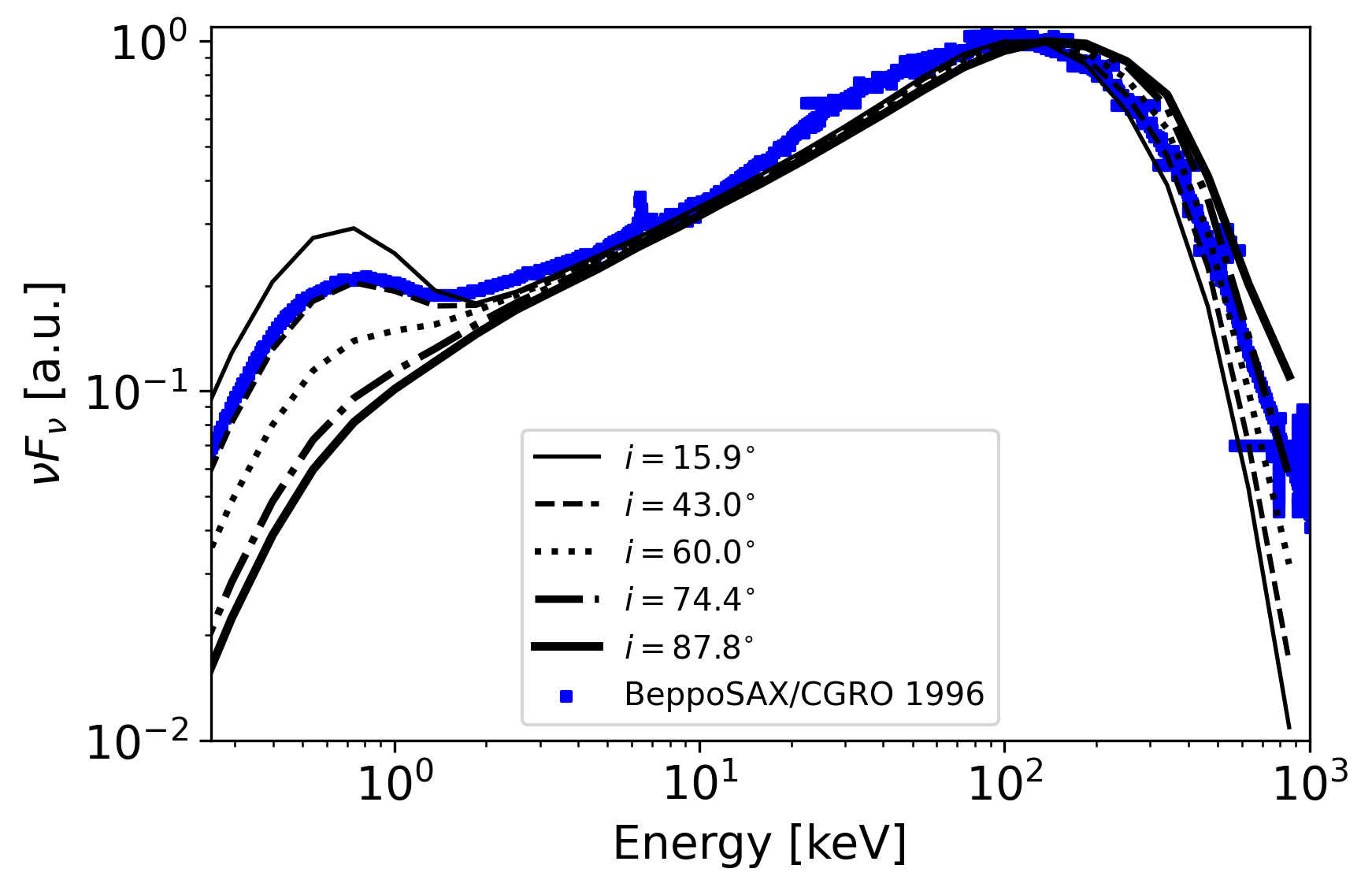}
    \caption{Spectral energy distribution for the fiducial simulation with $\tau=0.9$, $a_{min}=0.4$, $E_s=0.15$ keV, and the observed hard-state spectrum for comparison (blue), showing agreement with the observed spectral cutoff.}
    \label{fig:sed}
\end{figure}

In this section, we present results from the simplified toy model of a reconnection region. We assume the reconnection is in the equatorial plane of the black hole. We ignore the effects of the microscopic plasmoid geometry as well as Doppler frequency shifts from the
orbital motion of the reconnection zones and gravitational frequency shifts. The comparisons of the flux and polarization energy spectra with actual Cyg X-1 data should thus be interpreted with caution, as we neglect the actual geometry of the accretion flow as well as general relativistic effects. More detailed studies incorporating those effects will be discussed elsewhere (see Section \ref{discussion}). The spectral results for our simulation (Figure \ref{fig:sed}) are in broad agreement with observational data for the hard state of Cygnus X-1, in particular reproducing the 100~keV cutoff and photon index $\Gamma \sim 1.6-1.7$ typically attributed to thermal Comptonization \citep{1997MNRAS.288..958G}. We find the cutoff is well approximated by a reconnection layer with $\tau=0.9$ and $a_{min}=0.4$. Here, $\tau$ is chosen to reproduce the observed $\sim$ 100~keV cutoff, consistent with Beloborodov’s (\citeyear{2017ApJ...850..141B}) finding that $\tau_T \sim 1$ is the self-regulated value from pair balance, while $a_{min}$ sets the Compton amplification factor $A \sim 10$ needed to match the observed photon index, following Beloborodov’s framework where $a_{min}$ is a fit parameter for A. The deabsorbed spectrum shows a thermal peak of unscattered seed photons at low energies \citep{2003MNRAS.343L..84G}. We note that to match the location of this feature requires a seed temperature $E_s=0.15$ keV. Our power-law slightly underpredicts the flux from approximately 20 to 70 keV, which could be attributed to the contribution of photons reflected off the accretion disk that is not included in our model. At energies approaching $10^3$ keV, the low number of photons causes the statistics to be less reliable, and the model may over- or underproduce flux depending on the inclination.

\begin{figure}
    \centering
    \includegraphics[width=1.0\linewidth]{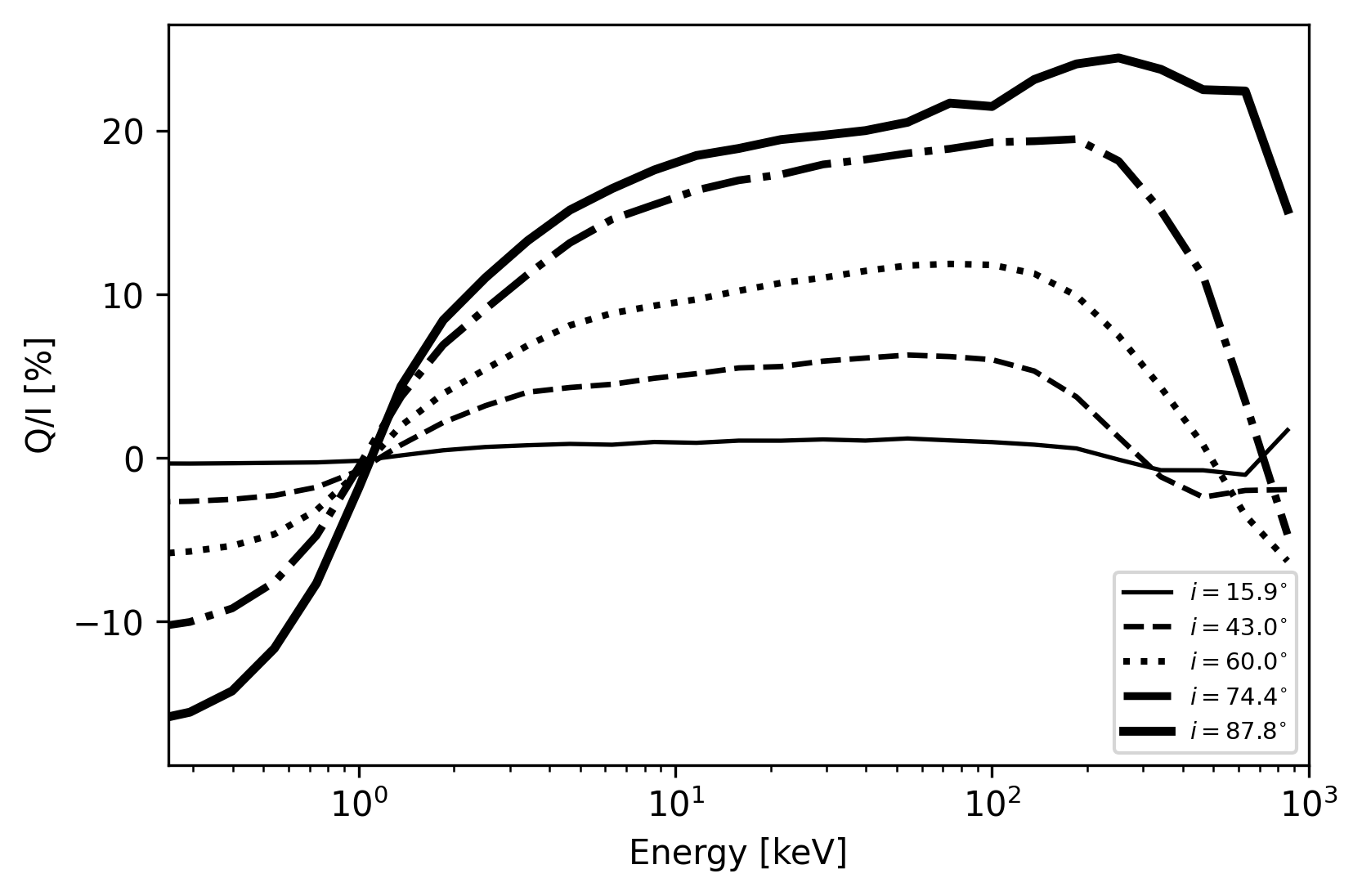}
    \caption{Polarization properties (Stokes $Q/I$) for $\tau=0.9$, $a_{\min}=0.4$, and $E_s=0.15$ keV. Positive $Q/I$-values correspond to electric field vectors being oriented preferentially perpendicular to the reconnection layer. }
    \label{fig:pol}
\end{figure}

\begin{figure}
    \centering
    \includegraphics[width=1.0\linewidth]{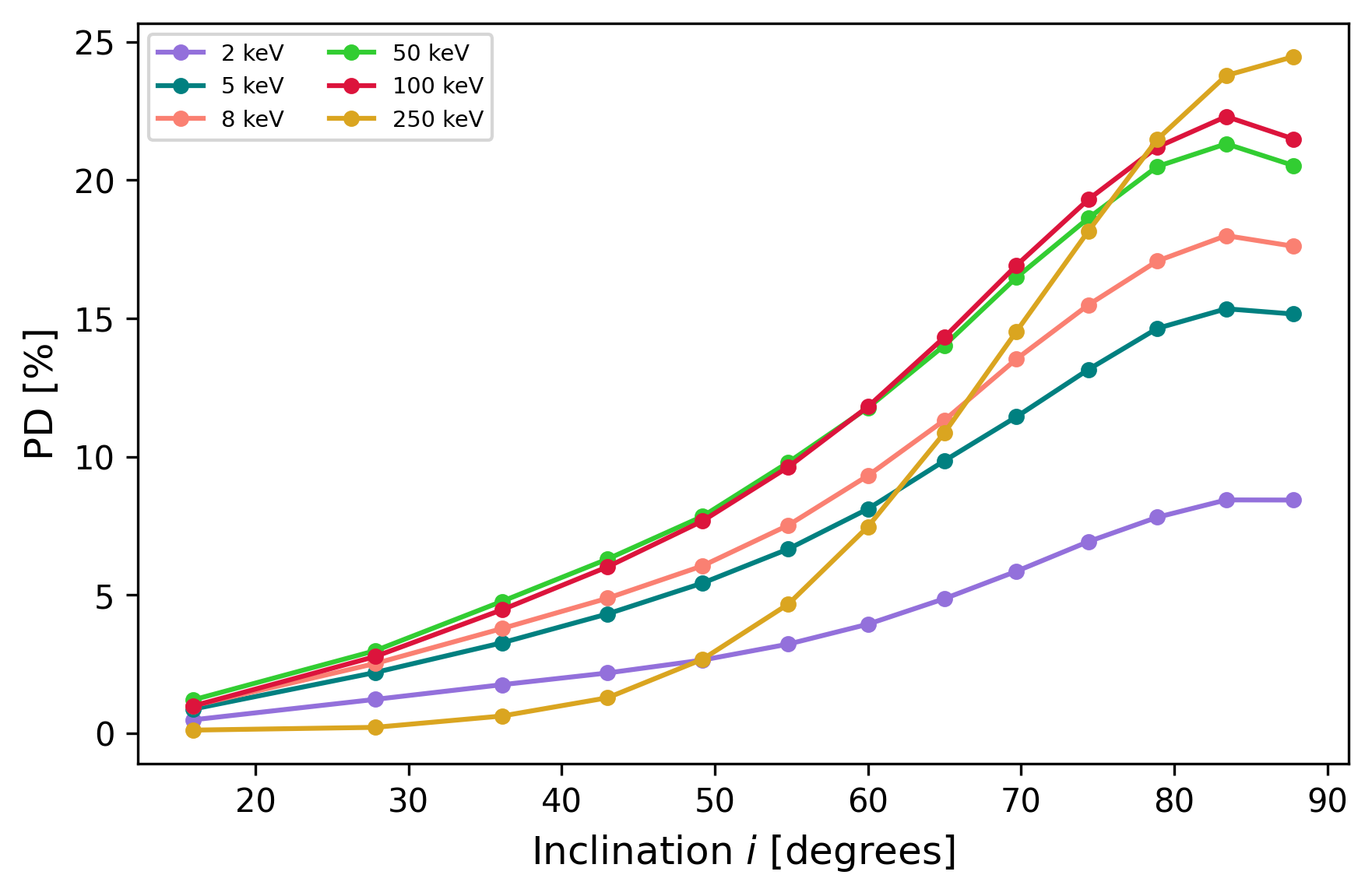}
    \caption{Dependence of PD on observer inclination for selected energies.
    We average over the orientations of the plasmoid motion, and the PD thus vanishes as the inclination goes to 0. It rises with inclination with the exception of the highest-inclination bin where the extreme geometries of scatterings lead to a small PD decrease.   All points are drawn at the centers of the energy bins. }
    \label{fig:incl}
\end{figure}

We observe a strong dependence of linear polarization on energy and observer inclination angle (Figure \ref{fig:pol}). We find that the Comptonized radiation at higher energies is polarized parallel to the layer normal, i.e., along the $z$-axis ($Q>0$ in our sign convention), while at low energies the polarization is perpendicular to the layer normal, i.e., in the plane of the layer ($Q<0$). The sign of $Q$ changes within a transition region around 1~keV. The maximum PD of $24\%$ at approximately 250~keV is achieved when viewing the layer nearly edge-on, and decreases for lower observer inclination. At the highest energies, the statistics are less reliable due to the relative lack of photons. These results are the average over azimuthal viewing angles. 
Figure \ref{fig:incl} shows the PD as a function of inclination in different energy bands.
We note rather high PDs exceeding those from the Comptonization in isotropic scattering atmospheres \citep{2010ApJ...712..908S,2017ApJ...850...14B,2022MNRAS.510.3674U,2025ApJ...995..124F}. 

\begin{figure*}[t]
\centering

\begin{subfigure}{0.48\textwidth}
\includegraphics[width=\linewidth]{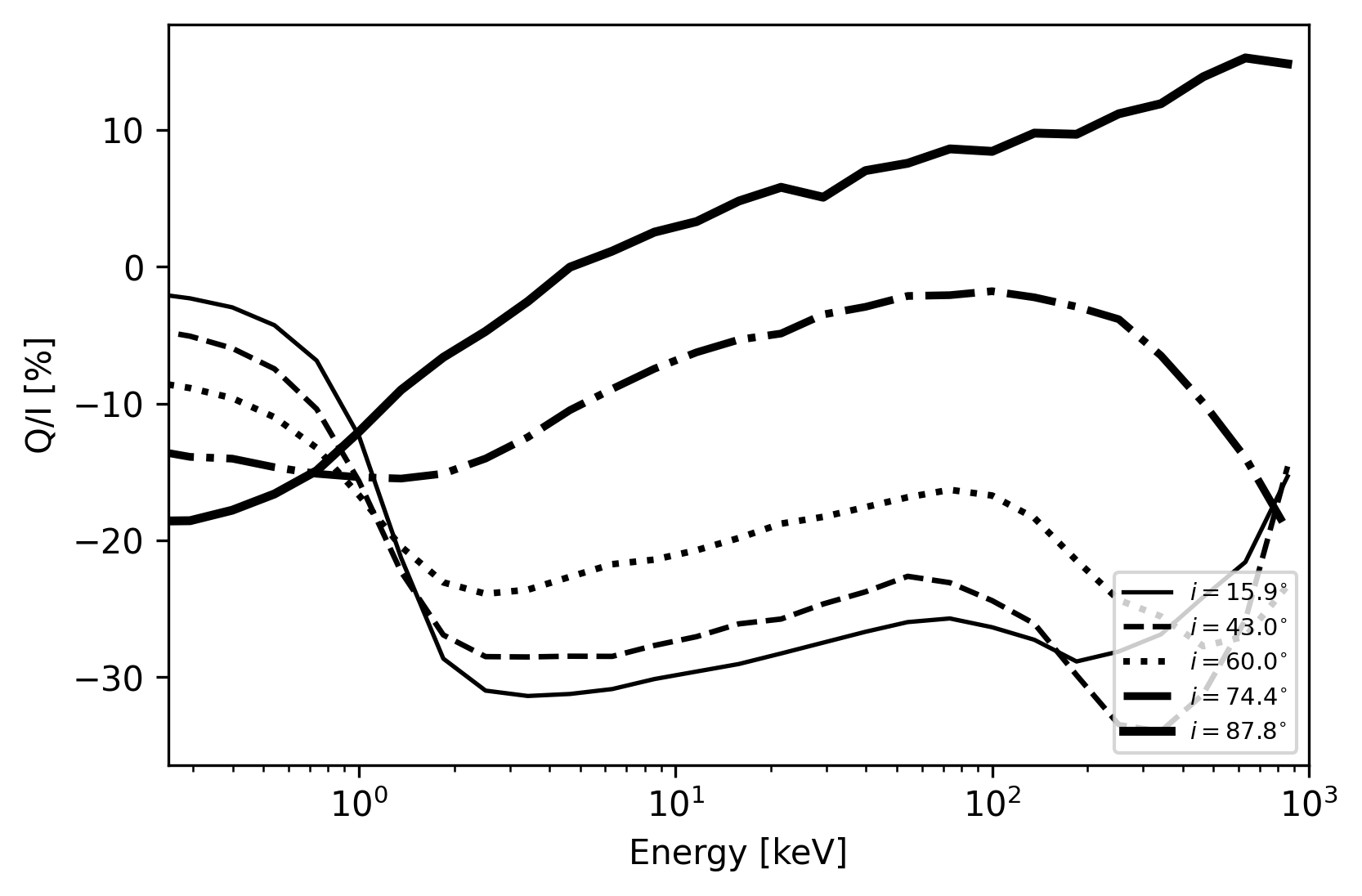}
\end{subfigure}
\hfill
\begin{subfigure}{0.48\textwidth}
\includegraphics[width=\linewidth]{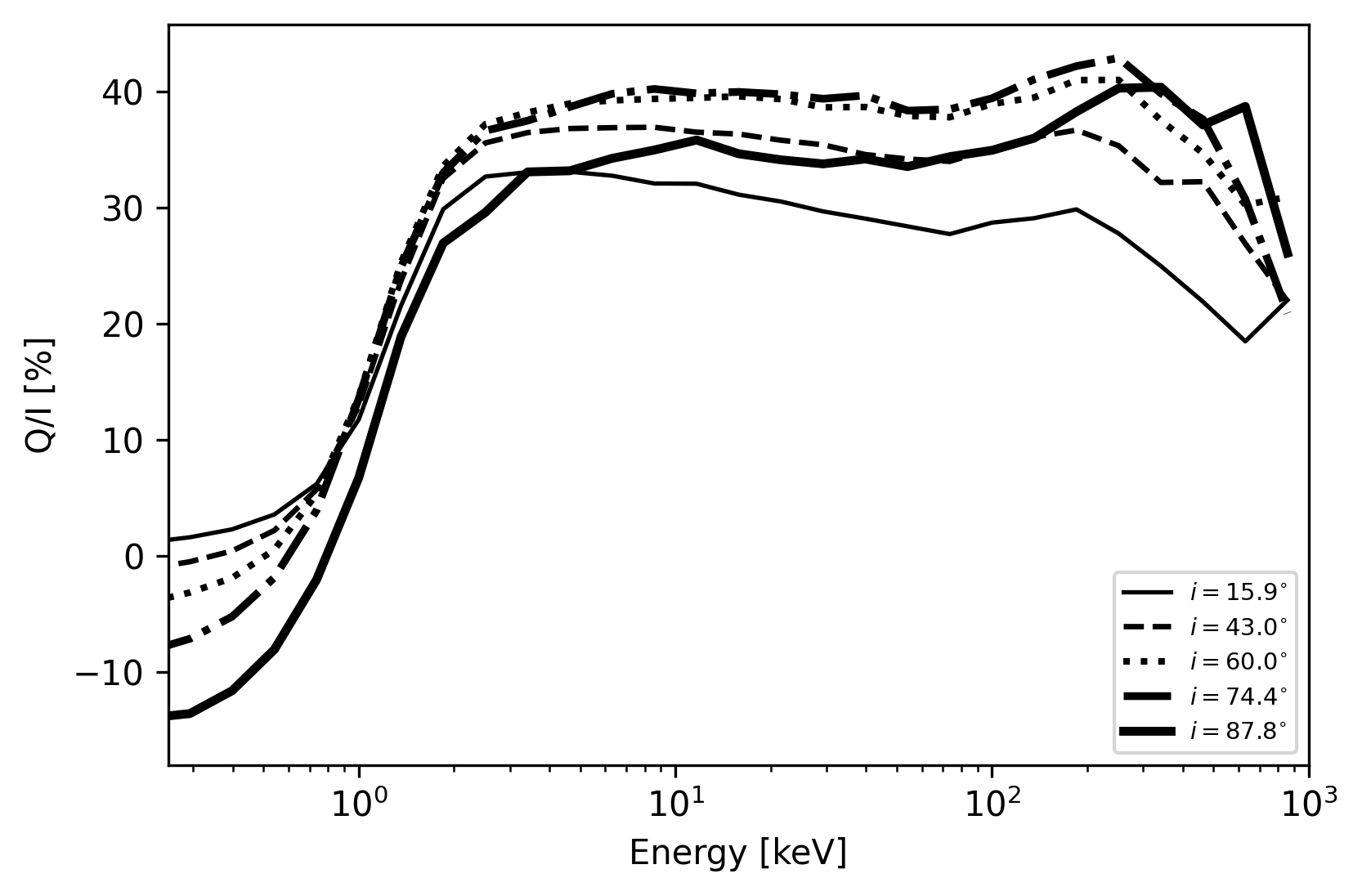}
\end{subfigure}

\caption{
Stokes $Q/I$ for 
viewing angles confined to $20^\circ$-wide azimuthal wedges. The left panel shows viewing directions centered on the positive $x$-axis, while the right panel shows those centered on the positive $y$-axis.}

\label{fig:testxy}
\end{figure*}

Next, we discuss the changes to polarization when the viewing angle is restricted to an azimuthal wedge of $20^{\circ}$ around the $x$- or $y$-axes. The results are displayed in Figure \ref{fig:testxy}. We find that the largest polarization perpendicular to the layer is produced for a viewing angle along y, reaching a maximum of 43.5\% for the bin centered at $i=69^{\circ}.7$. Conversely, the smallest perpendicular polarization and largest parallel polarization occurs when viewing along $x$. 

The results above can be understood by noting that the observed polarization direction, parallel or perpendicular to the reconnection layer, is determined by the orientations of the scattering planes in the rest frame of the scatterers. The PD and parallel or perpendicular orientation are Lorentz invariants. Furthermore, for fast scatterers the target photons always come in the reference frame of the scatterers from the direction into which they are moving. 
In Figure \ref{fig:pol}, the low-energy polarization is dominated by seed photons that have undergone few or no scatterings, retaining the polarization signature of Chandrasekhar's scattering atmosphere: polarization perpendicular to the normal ($Q<0$). At the lowest energies, the polarization is even stronger than that of the seed photons. This is caused by photons losing energy in scatterings. At higher energies, photons have been Comptonized primarily by the fast plasmoids moving along $\pm \hat x$. In the rest frame of these plasmoids, incoming photons arrive preferentially from the direction of plasmoid motion, so the scattering planes are oriented perpendicular to $\hat x$, producing polarization parallel to the layer normal ($Q>0$) in the observer frame. In contrast to the Chandrasekhar prediction of increasing PD toward the limb, always perpendicular to the disk normal and capped near 12\%, our model produces an energy-dependent sign flip in $Q$ and substantially higher peak PDs at near-edge-on inclinations.

The azimuthal viewing angle dependence in Figure \ref{fig:testxy} further illustrates the anisotropic nature of the scattering geometry. When the layer is viewed along the $x$-axis (left panel), the observer looks along the direction of plasmoid outflow. The relativistic boost of the plasmoids contributes little net polarization in this direction because the anisotropy of the scattered radiation is projected along the line of sight rather than onto the plane of the sky. As a result, $Q/I$ is reduced relative to the azimuthal average, and the plasmoid Comptonization produces a net polarization parallel to the reconnection sheet with $Q/I < 0$ for all inclinations except for viewers skimming the surface at $i=87^{\circ}.8$. 
When the layer is viewed along the $y$-axis (right panel), the observer is simultaneously perpendicular to the plasmoid outflow direction and to the plasma inflow direction. In this configuration, the scattering off plasmoids produces strong polarization perpendicular to the layer normal with $Q/I>0$. The maximum $Q/I$ substantially exceeds the prediction of the Chandrasekhar model and represents the strongest polarization signature achievable in this geometry.

\section{Discussion} \label{discussion}

The results of Section \ref{results} show that reconnection can produce strong polarization even when averaging over the azimuthal orientation of the plasmoid motion (Figure \ref{fig:pol}). The polarization is stronger than that from an isotropic scattering atmosphere as the plasmoid motion is confined to a plane. This result makes 
the mechanism an attractive candidate for explaining the strong hard-state X-ray polarization of sources like Cyg X-1 with PD$\,\sim\,4\%$ at $i\,\sim\,30^{\circ}$ source inclination
\citep[][]{2022Sci...378..650K} and IGR~J17091-3624 with a PD$\sim10$\%
\citep[][]{2025MNRAS.541.1774E,2025ApJ...989..165D}.

In the following, we discuss the expected net PDs for several locations of the reconnection layers in a black hole accretion flow. We consider two candidate sites for the reconnection layer: an equatorial current sheet in the innermost region of the accretion flow, and a jet sheath. These correspond to qualitatively different polarization predictions and can in principle be distinguished by current and future X-ray polarimetric observations.

\begin{figure}
    \centering
    \includegraphics[width=1.0\linewidth]{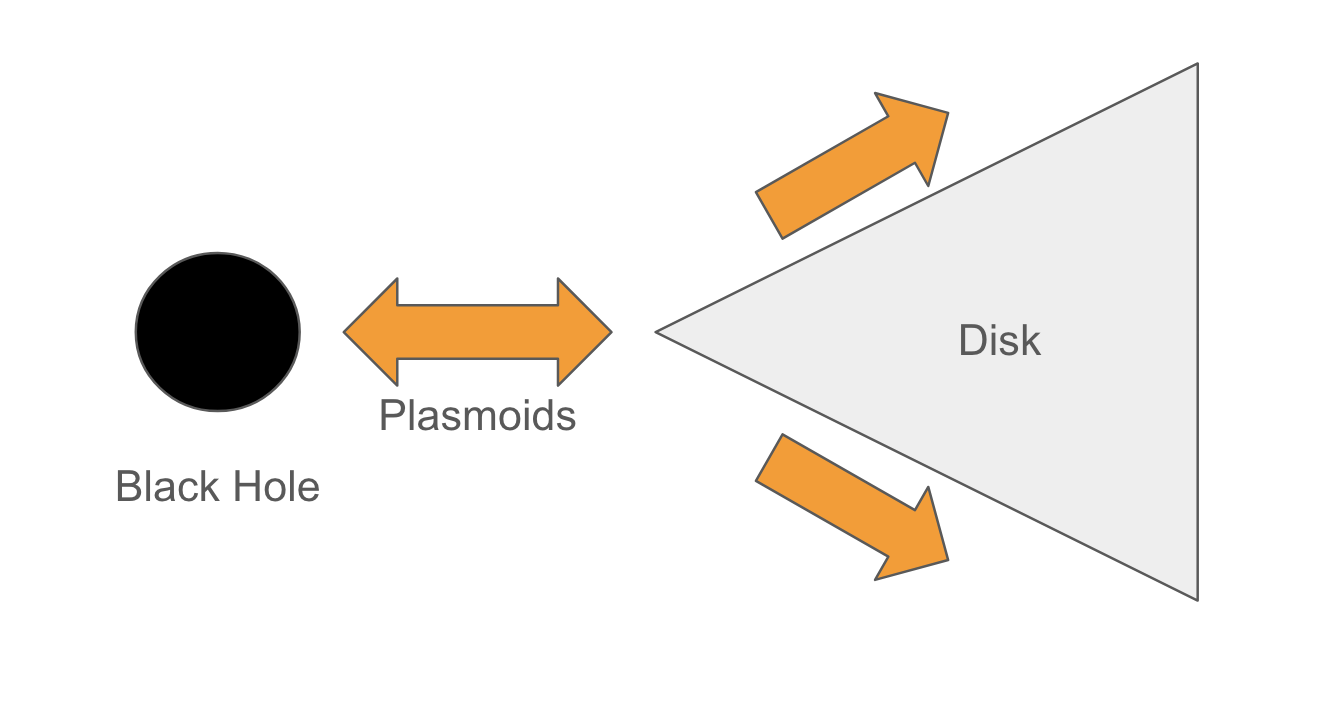}
    \caption{Proposed locations and directions of motion of  plasmoids for (i) an equatorial current sheet and (ii) jet sheaths or conical flows above and below the disk. We only show the radial and poloidal motion, but not the toroidal motion in and out of the plane of the image plane.}
    \label{fig:loc}
\end{figure}

The location and character of the reconnection region depends strongly on the
properties of the accretion flow. A schematic of possible locations within this configuration is provided in Figure \ref{fig:loc}. 
For a magnetically arrested disk (MAD) flow, ordered magnetic flux accumulation near the black hole can produce reconnection in the equatorial plane when fields of opposite orientation come into contact across the midplane \citep{2022ApJ...924L..32R, 2022ApJ...935L...1L}. Episodic flux eruptions in MAD disks are expected to drive energetic, large-scale reconnection events well suited to the cold plasmoid chain model. In a standard and normal evolution (SANE) accretion flow, the magnetic field is less ordered and reconnection events are likely to be more intermittent and lower in energy. 
The current observational state of Cyg X-1 remains ambiguous between these two scenarios. 

\begin{figure}
    \centering
    \includegraphics[width=1.0\linewidth]{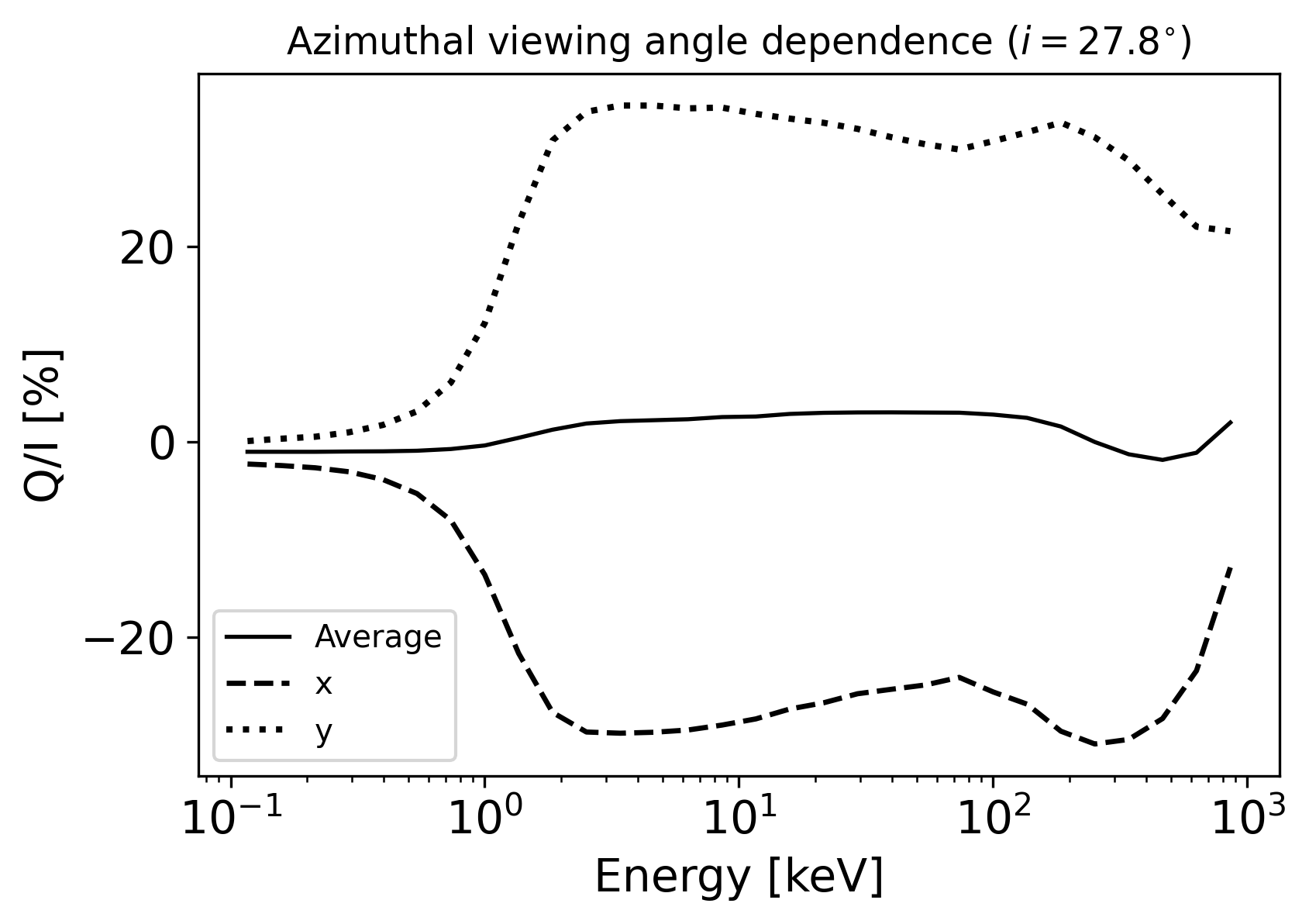}
    \caption{Polarization of the emission from an equatorial current when the plasmoids move perpendicular to the line of sight (along $x$), toward and away from the observer (along $y$), and when averaging over the equatorial plane for $i = 27^{\circ}.8$, corresponding to viewing a reconnection sheet in the equatorial plane of Cyg X-1.
    }
    \label{fig:28deg}
\end{figure}

An equatorial current sheet corresponds exactly to the midplane in our setup, and therefore viewing at approximately $i=28^{\circ}$ for Cyg X-1 is essentially equivalent to the $27^{\circ}.8$ inclination bin in our results (isolated in Figure \ref{fig:28deg}). In the azimuthal average, the maximum Q/I for this angle is only 3\%, substantially higher than the $\sim$1\%-2\% polarization from thermal Comptonization.
This scenario predicts a polarization direction perpendicular to the disk, parallel to the jet (if present).

\begin{figure}
    \centering
    \includegraphics[width=1.0\linewidth]{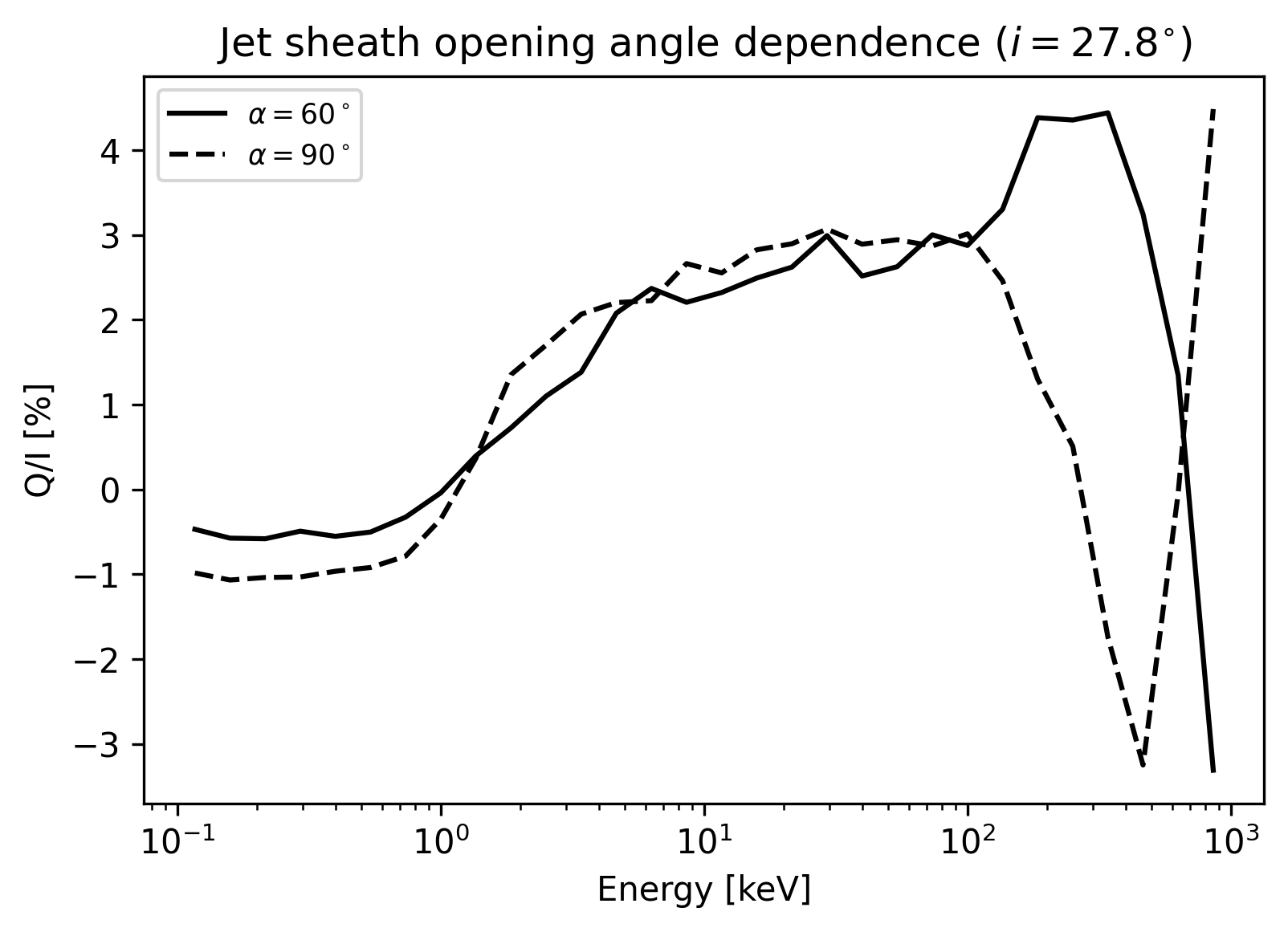}
    \caption{Dependence on simulated jet-sheath opening angle for $i = 27^{\circ}.8$.}
    \label{fig:cone}
\end{figure}

Alternatively, the Comptonizing region may be a conical or paraboloidal jet sheath \citep[e.g.,][]{2024MNRAS.528L.157D,2025ApJ...979..199S}. In this geometry, the reconnection layer wraps around the jet axis, and the azimuthal symmetry of the cone means that the polarization contributions from different segments of the sheet partially cancel upon integration over the whole emitting region. This averaging reduces the net PD relative to the equatorial plane geometry. However, if the reconnection layer is inclined relative to the $x-y$ plane, observer inclination angle relative to the layer normal would be larger than $i = 28^{\circ}$ (i.e., the observer is more aligned with the scattering plane). Given the strong dependence on inclination, and the relativistic beaming expected when viewing the plasmoid chain head-on, the PD could be still be large. To explore these effects, we simulate a range of angles $\alpha$ between the disk and the $z$-axis, corresponding to different jet-sheath opening angles. The results are generated in post-processing rather than changing the reconnection layer simulation. Specifically, after each photon is transported, the wavevector $\vec{k}$ and polarization vector $\vec{f}$ are rotated by $90^\circ-\alpha$ about the $y$-axis and then by a random number angle $\phi$ about the $z$-axis to average all segments of the cone. Polarization results for half-opening angles $\alpha = 60^{\circ}$, and $90^{\circ}$ are presented in Figure \ref{fig:cone}. At $\alpha = 90^{\circ}$, we recover our result from the equatorial current sheet configuration. 
For $\alpha = 60^{\circ}$, the net observed PD is slightly, but not substantially, higher.
It is likely that observations always average over the emission from reconnection regions seen from different angles. The results above show that snapshots of a single region could show a much stronger polarization (Figure \ref{fig:testxy}).    

The strong polarization of the plasmoid chain model makes it a candidate for explaining the {\it IXPE} observations of Cyg X-1 and IGR~J17091-3624.
For Cyg X-1, the model can explain the polarization degrees and angles---even though the observed PDs 
imply a slightly ($\sim5^{\circ}$) larger inclination than the binary inclination.
The $\sim$10\% PD of IGR~J17091-3624 requires a large inclination of about 70$^{\circ}$.

Note that we assumed here somewhat arbitrarily that the seed photons originate from the reconnection layer.
Seed photons originating from other sources, e.g., from the disk, would lead to different outcomes.  We also note that our time- averaged treatment neglects possible depolarization from unresolved temporal variability in the plasmoid chain, as well as from turbulence within the reconnection layer, which radiative PIC simulations show can be significant in magnetized coronae \citep[][]{2024PhRvL.132h5202G,2024NatCo..15.7026N}. We hope that the  results presented in this Letter will motivate additional studies of this promising mechanism.





\begin{acknowledgments}
The authors would like to thank Navin Sridhar, Bart Ripperda, Lorenzo Sironi, Daniel Gro{\v{s}}elj, Alex Chen, Yajie Yuan, Manel Errando, and John Mehlhaff for helpful discussions. The authors also thank the anonymous referee for constructive comments which improved this Letter. The authors acknowledge 
funding through NASA awards
80NSSC24K0205, 80NSSC24K1178, 80NSSC24K1819, and 80NSSC24K1749.
\end{acknowledgments}

\begin{contribution}
J.G. implemented the cold-chain Comptonization code in the Monte Carlo radiation transport code and wrote the major part of the Letter. K.H. and H.K. helped with the code and the scattering engine, and contributed to the interpretation of the results and the writing of the Letter.


\end{contribution}

%




\bibliography{sample701}{}
\bibliographystyle{aasjournalv7}



\end{document}